\documentclass[11pt]{article}
\usepackage{hyperref}
\begin{document}
\title{Complex Structure of Dynamic Stall on Wind Turbine Airfoils}
\author{Michael Hind, John Strike, Pourya Nikoueeyan, Andrew Magstadt, \\ 
       Ashli Babbitt, Phillip Davidson, and Jonathan Naughton \vspace{6pt} \\
 Mechanical Engineering Department and \\ Wind Energy Research Center, \\ University of Wyoming, Laramie, WY 82071, USA}
\maketitle
\begin{abstract}
Fluid dynamics video demonstrating the evolution of dynamic stall on a wind turbine blade.
\end{abstract}
Wind turbine blades frequently exhibit dynamic stall due to the unsteady flow they experience that cause changes in angle of attack the blade experiences.  Dynamic stall on pitching airfoils is dependent on the exact conditions the airfoil is experiencing as well as the shape of the airfoil.  As a result, it is important to be able to understand and classify the different types of dynamic stall observed.  To accomplish this goal, it is necessary to understand the characteristics of dynamic stall observed in both the flow field and the resulting pressure distribution.  In this way, the evolution of the flow field can be linked to the resulting forces and moments that are created.  Flow visualization is a powerful means portraying the dynamic stall process in order to better understand this complex flow.

A number of dynamic stall types have been observed in recent experimental measurements on wind turbine airfoils \cite{Naughton13}.  This fluid dynamics video considers a single case: a DU97-W-300 airfoil with a 10.2 cm chord $c$ oscillating in pitch $\alpha$,
\[ \alpha = \alpha_0 + \Delta \alpha \cos (2 \pi f t), \]
where $\alpha_0$ is the mean angle of attack, $\Delta \alpha$ is the pitch oscillation amplitude, $f$ is the frequency of oscillation, and $t$ is time.  For this case, $\alpha_0$ was 15$^\circ$, $\Delta \alpha$ was 10$^\circ$, and $f$ was 20 Hz.  The test was conducted at a free-stream velocity $U_\infty$ of 45 m/s yielding a Reynolds number of 220,000 and a reduced frequency ($k = \omega c / (2 U_\infty)$, where $\omega$ is the pitching frequency in radians per second) of 0.142.

It is clear from the video that combined surface and flow-field results are a substantial aid yo understanding the details of the dynamic stall process.  Results such as those shown in the video are also expected to be useful for validation of computational simulations of such flows.

\bibliographystyle{unsrt}

\begin{thebibliography}{10}

\bibitem{Naughton13}
J.~Naughton, J.~Strike, M.~Hind, A.~Magstadt, and A.~Babbitt.
\newblock Measurements of dynamic stall on the DU wind turbine airfoil series.
\newblock May 2013.
\newblock Presented at AHS Forum 69, Phoenix, AZ.

\end{thebibliography}

\end{document}